# Achieving thermodynamic consistency in a class of free-energy multiphase lattice Boltzmann models

Q. Li[*], Y. Yu, and R. Z. Huang

*School of Energy Science and Engineering, Central South University, Changsha 410083, China*

[*]Corresponding author: qingli@csu.edu.cn

**Abstract**

The free-energy lattice Boltzmann (LB) model is one of the major multiphase models in the LB community. The present study is focused on a class of free-energy LB models in which the divergence of thermodynamic pressure tensor or its equivalent form expressed by the chemical potential is incorporated into the LB equation via a forcing term. Although this class of free-energy LB models may be thermodynamically consistent at the continuum level, it suffers from thermodynamic inconsistency at the discrete lattice level owing to numerical errors [Guo *et al.*, Physical Review E **83**, 036707 (2011)]. The numerical error term mainly includes two parts, one comes from the discrete gradient operator and the other can be identified in a high-order Chapman-Enskog analysis. In this paper, we propose an improved scheme to eliminate the thermodynamic inconsistency of the aforementioned class of free-energy LB models. The improved scheme is constructed by modifying the equation of state of the standard LB equation, through which the discretization of $\nabla(\rho c_s^2)$ is no longer involved in the force calculation and then the numerical errors can be significantly reduced. Numerical simulations are subsequently performed to validate the proposed scheme. The numerical results show that the improved scheme is capable of eliminating the thermodynamic inconsistency and can significantly reduce the spurious currents in comparison with the standard forcing-based free-energy LB model.

# I. Introduction

The lattice Boltzmann (LB) method [1-5], which is a mesoscopic numerical approach originating from the lattice gas automaton (LGA) method [6], has been proven to be particularly suitable for studying multiphase and multicomponent systems [7,8] where the interfacial dynamics and phase transition are present. In the past three decades, significant progress has been made in this direction and a variety of multiphase LB models have been developed, such as the color-gradient LB model [9-11], the free-energy LB model [12-15], the pseudopotential LB model [16-21], and the phase-field LB model [22-24]. Although these models are devised from different points of view, they share the same feature, i.e., in these models the interface between different phases/components is represented by a diffuse interface. An important advantage of diffuse-interface models lies in that the motion of the liquid-gas interface does not need to be tracked explicitly [8].

Among these multiphase LB models, the free-energy model proposed by Swift *et al*. [12,13] was devised based on thermodynamic theory. They proposed to modify the second-order moment of the equilibrium density distribution function so as to include a non-ideal thermodynamic pressure tensor. Therefore in this model the phase separation is described by a non-ideal equation of state in thermodynamic theory such as the van der Waals equation of state. However, the original free-energy multiphase LB model was shown to break Galilean invariance due to some non-Navier-Stokes terms recovered in the momentum equation [25], which arises from incorporating the pressure tensor via the equilibrium density distribution function. To restore the Galilean invariance, several correction terms can be added to the equilibrium density distribution function [13-15,26].

Based on the consideration that the thermodynamics of a multiphase system can be equivalently taken into account through a forcing term, Wagner and Li [27] proposed a free-energy LB model that uses a forcing term to incorporate the divergence of the thermodynamic pressure tensor ($\nabla \cdot \mathbf{P}$) into the LB equation. In the meantime, a similar LB model was devised by Lee and Fischer [28] and they found that the spurious currents can be considerably reduced when the divergence of the pressure tensor is expressed by a chemical-potential form, i.e., $\nabla \cdot \mathbf{P} = \rho \nabla \mu_c$, where $\mu_c$ is the chemical potential. This class of free-energy LB models can be referred to as "forcing-based" free-energy models in comparison with the original free-energy LB model proposed by Swift *et al*. [12,13]. This forcing-based free-energy LB method has been recently extended to multiphase systems at large density ratios by Mazloomi M *et al*. [29] and Wen *et al*. [30].

Nevertheless, Guo *et al*. [31] found that, for the aforementioned class of free-energy LB models, the force balance condition does not hold at the discrete lattice level regardless of using the pressure-tensor form $\nabla \cdot \mathbf{P}$ or the chemical-potential form $\rho \nabla \mu_c$. Subsequently, Lou and Guo [32] showed that the force imbalance at the discrete level leads to thermodynamic inconsistency, i.e., the coexisting liquid and gas densities given by the forcing-based free-energy LB models gradually deviate from the results of the Maxwell construction when the reduced temperature decreases. As shown in Refs. [31,32], the force imbalance or the thermodynamic inconsistency of the forcing-based free-energy LB models is caused by the numerical errors at the discrete lattice level.

To reduce the effects of force imbalance, Lou and Guo proposed [32] a Lax-Wendroff propagation scheme for the forcing-based free-energy LB models. In a similar way, Qiao *et al*. [33] developed a forcing-based free-energy LB model using the Beam-Warming scheme. However, these two schemes are inconsistent with the philosophy of the standard streaming-collision procedure. The numerical errors are reduced by decreasing the Courant-Friedrichs-Lewy number, which leads to a much smaller time step than that of the standard LB method. As a result, the computational cost is significantly increased. Moreover, the numerical implementation of the Beam-Warming scheme involves not only the nearest-neighbor nodes but also the next-nearest-neighbor nodes.

In this paper, we aim to propose an alternative scheme that is still within the framework of the standard streaming-collision procedure for eliminating the thermodynamic inconsistency of the forcing-based free-energy LB models. Specifically, an improved scheme is constructed by modifying the equation of state of the standard LB equation, through which the discretization of $\nabla(\rho c_s^2)$ is no longer required in the force calculation and then the numerical errors can be significantly reduced. The rest of the present paper is organized as follows. In Sec. II, the forcing-based free-energy LB method is briefly introduced. The improved scheme is proposed in Sec. III and the numerical validation is provided in Sec. IV. Finally, a summary is given in Sec. V.

## II. Forcing-based free-energy LB method

### A. Basic formulations

For multiphase systems, the free energy functional can be expressed by [12,13,34-37]

$$\mathcal{F} = \int \psi(\rho, \nabla\rho) \mathrm{d}\Omega_V = \int \left( E_f(\rho) + \frac{\kappa}{2} |\nabla\rho|^2 \right) \mathrm{d}\Omega_V, \tag{1}$$

where $\Omega_V$ is the region of space occupied by the system, $\psi(\rho, \nabla\rho)$ is the free-energy density, $E_f(\rho)$ represents the bulk free-energy density, which leads to an equation of state that allows for the coexistence of liquid and gas phases, and $0.5\kappa|\nabla\rho|^2$ denotes the interfacial free-energy density, in which $\kappa$ is a positive constant. The chemical potential $\mu_c$ is defined as the variation of the free energy functional with respect to the density, i.e.,

$$\mu_c = \frac{\delta \mathcal{F}}{\delta \rho} = E'_f(\rho) - \kappa \nabla^2 \rho, \tag{2}$$

where $E'_f(\rho) = \mathrm{d}E_f(\rho)/\mathrm{d}\rho$. Then a non-local pressure can be defined as follows [38]:

$$p = \rho \mu_c - \psi(\rho, \nabla\rho). \tag{3}$$

Substituting Eq. (2) and the expression of $\psi(\rho, \nabla\rho)$ into Eq. (3) yields

$$p = p_{\mathrm{EOS}} - \kappa \rho \nabla^2 \rho - \frac{\kappa}{2} |\nabla\rho|^2, \tag{4}$$

where $p_{\mathrm{EOS}} = \rho E'_f(\rho) - E_f(\rho)$ is the non-ideal equation of state. Correspondingly, the thermodynamic pressure tensor is defined as [38-40]

$$\mathbf{P} = p\mathbf{I} + \frac{\partial \psi(\rho, \nabla\rho)}{\partial(\nabla\rho)} \nabla\rho = p\mathbf{I} + \kappa \nabla\rho \nabla\rho, \tag{5}$$

where $\mathbf{I}$ is the unit tensor. After some standard algebra, the following equation can be obtained [8,28]

$$\nabla \cdot \mathbf{P} = \rho \nabla \mu_c. \tag{6}$$

The left-hand side of Eq. (6) is usually referred to as the pressure-tensor form, whereas the right-hand side is referred to as the chemical-potential form.

For the forcing-based free-energy LB method, the thermodynamics of a multiphase system is taken into account through a forcing term [27,28]. Without loss of generality, in the present study we adopt the LB equation with a multiple-relaxation-time (MRT) collision operator [41,42] based on the consideration that the single-relaxation-time (SRT) collision operator [43] is a special case of the MRT collision operator. Generally, the MRT-LB equation can be written as follows [44,45]:

$$f_\alpha(\mathbf{x} + \mathbf{e}_\alpha \delta_t, t + \delta_t) = f_\alpha(\mathbf{x}, t) - \overline{\Lambda}_{\alpha\beta}(f_\beta - f_\beta^{eq})\big|_{(\mathbf{x},t)} + \delta_t (G_\alpha - 0.5\overline{\Lambda}_{\alpha\beta} G_\beta)\big|_{(\mathbf{x},t)}, \tag{7}$$

where $f_\alpha$ is the density distribution function, $f_\alpha^{eq}$ is the equilibrium density distribution function, $\mathbf{x}$ is the spatial position, $t$ is the time, $\delta_t$ is the time step, $\mathbf{e}_\alpha$ is the discrete velocity in the $\alpha$ th direction, $G_\alpha$ is the forcing term in the discrete velocity space, and $\overline{\Lambda}_{\alpha\beta} = (\mathbf{M}^{-1}\Lambda\mathbf{M})_{\alpha\beta}$ is the collision operator, in which $\mathbf{M}$ is a transformation matrix and $\Lambda$ is a diagonal matrix for the relaxation times [44,45]. For the two-dimensional nine-velocity (D2Q9) lattice, the relaxation matrix is given by

$$\Lambda = \text{diag}\left(\tau_c^{-1}, \tau_e^{-1}, \tau_\varsigma^{-1}, \tau_c^{-1}, \tau_q^{-1}, \tau_c^{-1}, \tau_q^{-1}, \tau_v^{-1}, \tau_v^{-1}\right), \tag{8}$$

where $\tau_c = 1$ is the non-dimensional relaxation time related the conserved moments, $\tau_\varsigma$ and $\tau_q$ are free parameters, while $\tau_e$ and $\tau_v$ are the non-dimensional relaxation times determining the bulk and shear viscosities, respectively, i.e., $\eta_B = \rho c_s^2 (\tau_e - 0.5)\delta_t$ and $\eta = \rho c_s^2 (\tau_v - 0.5)\delta_t$.

Through the transformation matrix $\mathbf{M}$, the right-hand side of Eq. (7), namely the collision step, can be implemented in the moment space:

$$\mathbf{m}^* = \mathbf{m} - \Lambda(\mathbf{m} - \mathbf{m}^{eq}) + \delta_t \left(\mathbf{I} - \frac{\Lambda}{2}\right)\mathbf{S}, \tag{9}$$

where $\mathbf{m} = \mathbf{M}\mathbf{f}$, $\mathbf{m}^{eq} = \mathbf{M}\mathbf{f}^{eq}$, and $\mathbf{S} = \mathbf{M}\mathbf{G}$ is the forcing term in the moment space [5]. Then the streaming step can be implemented as follows:

$$f_\alpha(\mathbf{x} + \mathbf{e}_\alpha \delta_t, t + \delta_t) = f_\alpha^*(\mathbf{x}, t), \tag{10}$$

where $\mathbf{f}^* = \mathbf{M}^{-1}\mathbf{m}^*$ and $\mathbf{M}^{-1}$ is the inverse matrix of the transformation matrix. For the D2Q9 lattice, the standard equilibria $\mathbf{m}^{eq}$ are given by

$$\mathbf{m}^{eq} = \rho\left(1,\ -2+3u^2,\ 1-3u^2,\ u_x,\ -u_x,\ u_y,\ -u_y,\ u_x^2-u_y^2,\ u_x u_y\right)^{\mathrm{T}}, \tag{11}$$

where $u^2 = u_x^2 + u_y^2$. The macroscopic density and velocity are calculated by

$$\rho = \sum_\alpha f_\alpha,\quad \rho\mathbf{u} = \sum_\alpha \mathbf{e}_\alpha f_\alpha + \frac{\delta_t}{2}\mathbf{F}, \tag{12}$$

where $\mathbf{F}$ is the force exerted on the system. For the forcing-based free-energy LB method, the force is defined as follows [27,28]:

$$\mathbf{F} = \nabla\left(\rho c_s^2\right) - \nabla\cdot\mathbf{P}\quad \text{or}\quad \mathbf{F} = \nabla\left(\rho c_s^2\right) - \rho\nabla\mu_c, \tag{13}$$

where $c_s^2 = 1/3$. The left one is the pressure-tensor form, whereas the right one is the chemical-potential form. The forcing term $G_\alpha$ in Eq. (7) is given by [46]

$$G_\alpha = \omega_\alpha\left[\frac{\mathbf{e}_\alpha - \mathbf{u}}{c_s^2} + \frac{(\mathbf{e}_\alpha\cdot\mathbf{u})}{c_s^4}\mathbf{e}_\alpha\right]\cdot\mathbf{F}, \tag{14}$$

where $\omega_\alpha$ are the weights. For the D2Q9 lattice, $\omega_\alpha$ are given by $\omega_0 = 4/9$, $\omega_{1-4} = 1/9$, and $\omega_{5-8} = 1/36$. With the aid of Eq. (14), the forcing term in the moment space can be obtained via $\mathbf{S} = \mathbf{MG}$, in which $\mathbf{G} = (G_0, G_1, \cdots, G_8)^{\mathrm{T}}$.

### B. Numerical error term at the discrete level

The macroscopic equations recovered from the forcing-based free-energy LB method can be derived by the Taylor expansion analysis [47] or the Chapman-Enskog analysis [48]. As argued by Wagner [49], a second-order analysis is inadequate for multiphase LB models because higher-order terms are ignored, which may be necessary to achieve thermodynamic consistency. Through a third-order Chapman-Enskog analysis [50], the following macroscopic equations can be obtained:

$$\frac{\partial \rho}{\partial t} + \nabla\cdot(\rho\mathbf{u}) = 0, \tag{15}$$

$$\frac{\partial(\rho\mathbf{u})}{\partial t} + \nabla\cdot(\rho\mathbf{u}\mathbf{u}) = -\nabla\left(\rho c_s^2\right) + \mathbf{F} + \nabla\cdot\mathbf{\Pi} + \frac{\delta_t^2}{12}\nabla\cdot\nabla\mathbf{F}, \tag{16}$$

where $\mathbf{\Pi}$ is the viscous stress tensor. The last term on the right-hand side of Eq. (16) is a high-order term that cannot be identified by a second-order Chapman-Enskog analysis.

Meanwhile, in numerical implementation the gradient terms are usually evaluated by the following isotropic finite-difference scheme:

$$\nabla \phi^{(\text{discrete})} = \frac{1}{c_s^2 \delta_t} \sum_\alpha \omega_\alpha \phi(\mathbf{x} + \mathbf{e}_\alpha \delta_t) \mathbf{e}_\alpha . \tag{17}$$

According to the Taylor series expansion, we can obtain

$$\phi(\mathbf{x} + \mathbf{e}_\alpha \delta_t) = \phi(\mathbf{x}) + e_{\alpha k} \delta_t \partial_k \phi(\mathbf{x}) + \frac{\delta_t^2}{2} e_{\alpha k} e_{\alpha l} \partial_k \partial_l \phi(\mathbf{x}) + \frac{\delta_t^3}{6} e_{\alpha k} e_{\alpha l} e_{\alpha m} \partial_k \partial_l \partial_m \phi(\mathbf{x}) + \cdots . \tag{18}$$

Substituting Eq. (18) into Eq. (17) gives

$$\nabla \phi^{(\text{discrete})} = \nabla \phi + \frac{\delta_t^2}{6} \nabla \nabla^2 \phi + \cdots . \tag{19}$$

In the LB method we usually adopt $\delta_t = c = 1$, where $c$ is the lattice constant. Hence the force at the discrete level can be expressed via

$$\mathbf{F}^{(\text{discrete})} = \mathbf{F} + \frac{1}{6} \nabla(\nabla \cdot \mathbf{F}) + \cdots . \tag{20}$$

Combining Eq. (16) with Eq. (20), we can obtain the following numerical error term:

$$\mathcal{E} = \frac{1}{6} \nabla(\nabla \cdot \mathbf{F}) + \frac{1}{12} \nabla \cdot \nabla \mathbf{F} + O(\partial^4) . \tag{21}$$

At the thermodynamically equilibrium state, the chemical potential is constant everywhere, including the liquid-gas interface where the density varies. Accordingly, the major numerical error term is given by

$$\mathcal{E}_n = \frac{c_s^2}{6} \nabla(\nabla \cdot \nabla \rho) + \frac{c_s^2}{12} \nabla \cdot \nabla \nabla \rho . \tag{22}$$

Since $\partial_j \partial_i \partial_i \rho = \partial_i \partial_i \partial_j \rho$, we can rewrite Eq. (22) as

$$\mathcal{E}_n = \frac{c_s^2}{4} \nabla(\nabla^2 \rho) = \frac{1}{12} \nabla(\nabla^2 \rho) . \tag{23}$$

Note that $\nabla(\nabla^2 \rho) = \nabla \cdot [(\nabla^2 \rho)\mathbf{I}]$. Consequently, the isotropic part of the thermodynamic pressure tensor is changed by the numerical error term, which leads to the thermodynamic inconsistency of the forcing-based free-energy LB models.

### III. Alternative scheme for achieving thermodynamic consistency

In this section, an improved scheme that is still within the framework of the standard streaming-collision procedure is proposed to eliminate the thermodynamic inconsistency of the forcing-based free-energy LB models. In fact, the appearance of $\nabla(\rho c_s^2)$ in the force given by Eq. (13) is attributed to the equation of state $p = \rho c_s^2$ produced by the standard LB equation, which yields the term $-\nabla(\rho c_s^2)$ on the right-hand side of Eq. (16). Obviously, when such an equation of state is modified as $p = p_m$, the term $\nabla(\rho c_s^2)$ in Eq. (13) can be replaced by $\nabla p_m$. In the present study we utilize $p_m = (1 + \mu_c)/3$. Then the chemical-potential form of the force is given by

$$\mathbf{F} = \nabla p_m - \rho \nabla \mu_c = \left(\frac{1}{3} - \rho\right) \nabla \mu_c. \tag{24}$$

In order to modify the equation of state produced by the standard LB equation, the second-order moment of the equilibrium density distribution function can be defined as follows:

$$\sum_\alpha e_{\alpha i} e_{\alpha j} f_\alpha^{eq} = p_m \delta_{ij} + \rho u_i u_j, \tag{25}$$

where $\delta_{ij}$ is the Kronecker delta.

Nevertheless, when Eq. (25) is implemented without other changes, the LB model will suffer from the lack of Galilean invariance owing to some non-Navier-Stokes terms recovered in the macroscopic momentum equation. Such an issue can be found in the original free-energy LB model and the color-gradient LB models with variable density ratios [10,51]. Several approaches have been devised in the literature to address this issue [13,14,26,52]. In the present study, we adopt the approach proposed by Li *et al.* [52], which has been recently applied to eliminate the error terms of the color-gradient LB models [53,54]. Following this approach, the third-order moment of the equilibrium density distribution function is given by

$$\sum_\alpha e_{\alpha i} e_{\alpha j} e_{\alpha k} f_\alpha^{eq} = \begin{cases} \rho c_s^2 (u_i \delta_{jk} + u_j \delta_{ik} + u_k \delta_{ij}), & \text{if } i = j = k, \\ p_m (u_i \delta_{jk} + u_j \delta_{ik} + u_k \delta_{ij}), & \text{others}. \end{cases} \tag{26}$$

According to Eqs. (25) and (26), the equilibria $\mathbf{m}^{eq}$ in the moment space are now given by

$$\mathbf{m}^{eq} = \rho\left(1,\ -4 + 3u^2 + 2\gamma,\ 4 - 3u^2 - 3\gamma,\ u_x,\ (\gamma - 2)u_x,\ u_y,\ (\gamma - 2)u_y,\ u_x^2 - u_y^2,\ u_x u_y\right)^T, \tag{27}$$

where $\gamma = 3p_m/\rho$.

From Eq. (26) it can be found that the diagonal elements ($i = j = k$) of the third-order moment deviate from the require relationship. As a result, there are still a couple of error terms that should be removed by introducing a correction term into the LB equation, i.e.,

$$\mathbf{m}^* = \mathbf{m} - \mathbf{\Lambda}(\mathbf{m} - \mathbf{m}^{eq}) + \delta_t\left(\mathbf{I} - \frac{\mathbf{\Lambda}}{2}\right)(\mathbf{S} + \mathbf{C}), \tag{28}$$

in which the correction term $\mathbf{C}$ is given by $\mathbf{C} = (0, 9C_1, 0, 0, 0, 0, 0, 3C_7, 0)^{\mathrm{T}}$ with

$$C_1 = \partial_x(\varphi u_x) + \partial_y(\varphi u_y), \quad C_7 = \partial_x(\varphi u_x) - \partial_y(\varphi u_y), \tag{29}$$

where $\varphi = (\rho c_s^2 - p_m)$. The dynamic shear viscosity is now given by $\eta = p_m(\tau_v - 0.5)\delta_t$.

To sum up, Eqs. (24), (27), and (28) constitute the improved scheme for eliminating the thermodynamic inconsistency of the forcing-based free-energy LB models. Compared with the standard forcing-based free energy LB model, the present scheme introduces some additional computations, namely the discretization of $\partial_x(\varphi u_x)$ and $\partial_y(\varphi u_y)$. But it should also be noted that the discretization of $\nabla(\rho c_s^2)$, which is required in the standard forcing-based free energy LB model, is no longer needed when using the improved scheme.

## IV. Numerical validation

In the preceding section, we have proposed an improved scheme for eliminating the thermodynamic inconsistency of the forcing-based free-energy LB models. In this section numerical simulations are carried out to validate the proposed scheme.

### A. Flat interface

Firstly, the test of one-dimensional flat interfaces is considered. The grid system is taken as $L_x \times L_y = 100 \times 100$ with the periodic boundary condition being applied in the $x$ and $y$ directions. Initially, the flat interfaces are placed at $x = 0.25L_x$ and $x = 0.75L_x$, namely the central region is filled with the liquid phase, while the rest is occupied by the gas phase. The van der Waals equation of state is employed [12,13,55], which corresponds to the bulk free-energy density $E_f(\rho) = \rho RT \ln[\rho/(1-b\rho)] - a\rho^2$. Then the following chemical potential can be obtained according to Eq. (2):

$$\mu_c = RT\left[\ln\left(\frac{\rho}{1-b\rho}\right) + \frac{1}{1-b\rho}\right] - 2a\rho - \kappa\nabla^2\rho, \tag{30}$$

where $a$ is the attraction parameter, $b$ is the repulsion parameter, and $R$ is the gas constant. In our simulations, the parameters are chosen as follows: $a = 9/392$, $b = 2/21$, $R = 1$, and $\kappa = 0.02$. The critical density and temperature are given by $\rho_c = 1/(3b) = 3.5$ and $T_c = 8a/(27Rb) = 1/14$.

Figure 1 displays the numerical coexistence curves predicted by the standard forcing-based free-energy LB model and the free-energy LB model using the improved scheme in the cases of $\nu = \eta/\rho = 0.15$ and $0.03$, where $\nu$ is the kinematic viscosity. For comparison, the analytical solution given by the Maxwell construction is also shown there. From the figure it can be clearly seen that the numerical coexistence curve given by the standard forcing-based free-energy LB model gradually deviates from the analytical one as the reduced temperature decreases. In contrast, the numerical coexistence curve yielded by the improved scheme agrees well with the analytical one. Particularly, no deviations are observed between the cases of $\nu = 0.15$ and $\nu = 0.03$, which indicates the thermodynamic consistency of the improved scheme is independent of the viscosity.

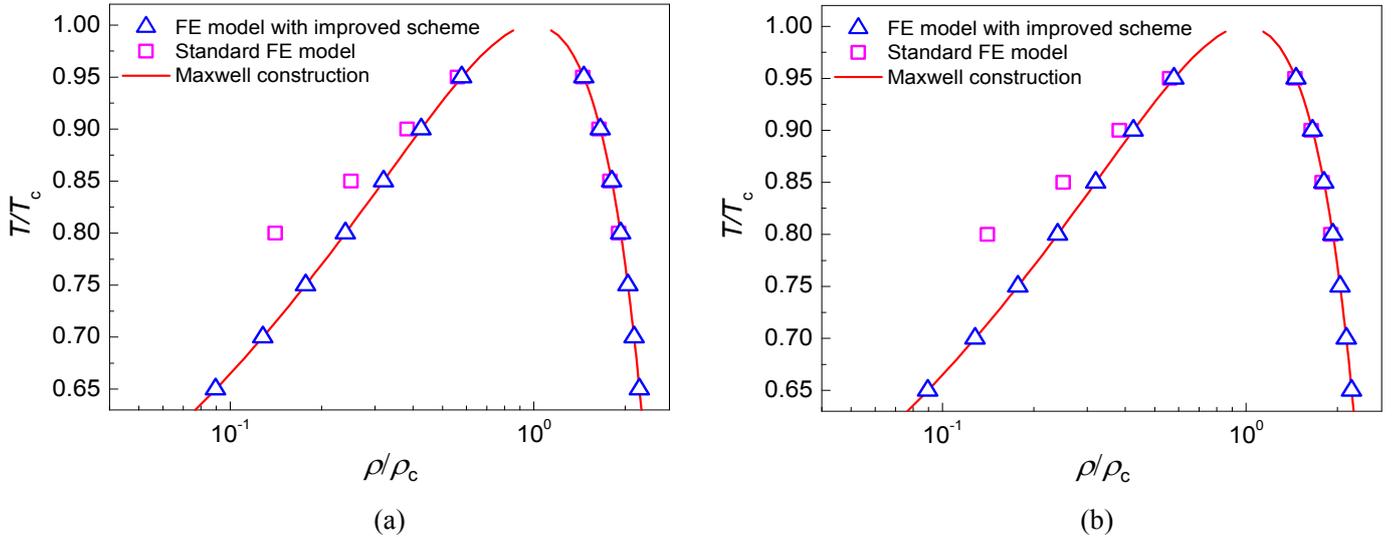

(a)        (b)

**FIG. 1**. Comparison of the numerical coexistence curves with the analytical solution given by the Maxwell construction. The kinematic viscosity is given by (a) $v = 0.15$ and (b) $0.03$.

At the thermodynamically equilibrium state, the chemical potential is constant everywhere. However, the standard forcing-based free-energy LB model results in a non-constant chemical potential due to the force imbalance at the discrete lattice level [32], as shown in Fig. 2, where the chemical potential profiles obtained by the standard forcing-based free-energy LB model and the free-energy LB model using the improved scheme are compared at $T = 0.8T_c$ with $v = 0.15$. It can be seen that the chemical potential predicted by the standard forcing-based free-energy LB model varies significantly across the interfaces. Contrarily, the free-energy LB model using the improved scheme produces a rigorously uniform chemical potential ($\mu_c \approx 0.018302$) in the whole computational domain. Moreover, a numerical comparison between the improved scheme and the Lax-Wendroff propagation scheme formulated by Lou and Guo [32] can be found in the appendix A, from which it can be seen that the improved scheme yields a uniform chemical potential in all the investigated cases, whereas the Lax-Wendroff propagation scheme leads to a fluctuating chemical potential with variations across the interfaces.

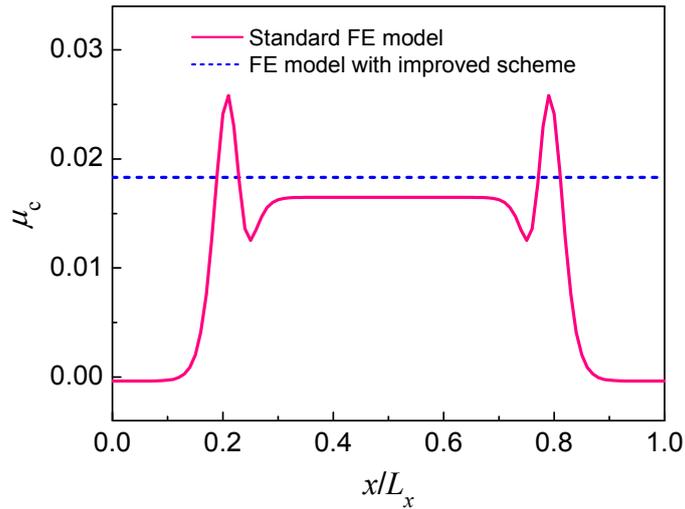

**FIG. 2**. Comparison of the chemical potential profiles predicted by the standard forcing-based free-energy LB model and the free-energy LB model using the improved scheme at $T = 0.8T_c$.

Furthermore, it is noticed that the numerical error term given by Eq. (23) not only alters the coexisting liquid and gas densities but also affects the interface thickness. Figure 3 displays the density profiles obtained by the standard forcing-based free-energy LB model and the free-energy LB model using the improved scheme at $T = 0.8T_c$ with $v = 0.15$. From the figure we can see that the liquid-gas interface predicted by the standard forcing-based free-energy LB model is much thicker than that given by the improved scheme. Such a phenomenon indicates that the major numerical error term described by Eq. (23) serves as a numerical dissipation term for the forcing-based free-energy LB models, which smoothes the liquid-gas interface. Here it is also worth mentioning that the interface in the LB simulations usually becomes thinner with the decrease of the reduced temperature (see Fig. 3 in Ref. [19] for details), but it can be widened by adjusting the parameter $a$ in the non-ideal equation of state.

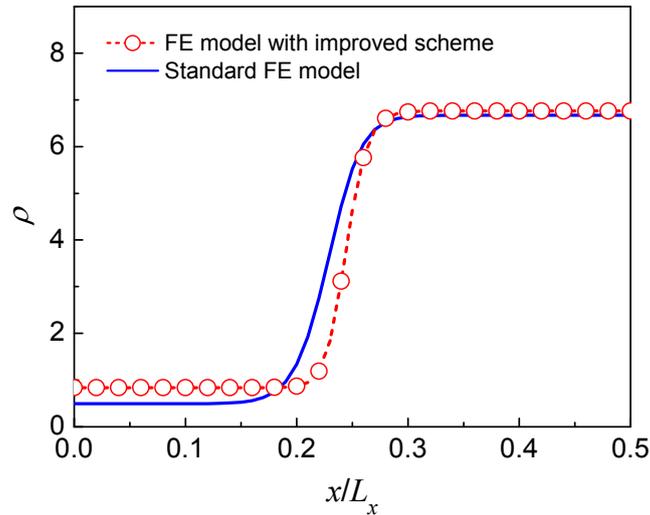

**FIG. 3**. Comparison of the density profiles obtained by the standard forcing-based free-energy LB model and the free-energy LB model using the improved scheme at $T = 0.8T_c$.

### B. Circular droplet

In this subsection, numerical simulations are performed for the problem of two-dimensional circular droplet. In this test, a liquid droplet of radius $r$ is placed at the center of a square domain and the rest of the domain is filled with the gas phase. The grid system is chosen as $L_x \times L_y = 120 \times 120$ with the periodic boundary condition being applied in the $x$ and $y$ directions. The density field is initialized as follows:

$$\rho(x,y) = \frac{\rho_l + \rho_g}{2} - \frac{\rho_l - \rho_g}{2} \tanh\left[\frac{2(r_0 - r)}{W}\right], \tag{31}$$

where $\rho_l$ and $\rho_g$ are the densities of the liquid and gas phases, respectively, $W = 5$ is the initial interface thickness, and $r_0 = \sqrt{(x - x_0)^2 + (y - y_0)^2}$, in which $(x_0, y_0)$ is the center of the computational domain.

For circular droplets, the liquid and gas densities at the equilibrium state can be obtained according to the non-ideal equation of state and the Young-Laplace's law [32]. The equilibrium liquid and gas densities of flat surfaces correspond to the case of a circular droplet with the radius $r \to \infty$. As the droplet size decreases, both the liquid and gas densities increase above the respective flat-interface values according to the Young-Laplace's law, which gives the following expressions for the pressures inside and outside of the circular droplet, respectively [56]:

$$p_l = p_l^{\text{sat}} + \frac{\rho_l^{\text{sat}}}{\rho_l^{\text{sat}} - \rho_g^{\text{sat}}} \frac{\vartheta}{r}, \quad p_g = p_g^{\text{sat}} + \frac{\rho_g^{\text{sat}}}{\rho_l^{\text{sat}} - \rho_g^{\text{sat}}} \frac{\vartheta}{r}, \tag{32}$$

where $p_l$ and $p_g$ are the pressures of the liquid and gas phases, respectively, $\vartheta$ is the surface tension, and the superscript sat denotes the properties given by the Maxwell construction. When the surface tension $\vartheta$ and the droplet radius $r$ are known, the pressures $p_l$ and $p_g$ can be determined at a given reduced temperature ($T/T_c$). Then the analytical liquid and gas densities can be theoretically obtained by solving $p_l = p_{\text{EOS}}(\rho_l)$ and $p_g = p_{\text{EOS}}(\rho_g)$, respectively. In the present test, the parameters $a$, $b$, $R$, and $\kappa$ are the same as those used in the previous section.

The numerical liquid and gas densities obtained by the standard forcing-based free-energy LB model and the free-energy LB model using the improved scheme are displayed in Figs. 4 and 5 for the cases of $T/T_c = 0.8$ and $0.7$, respectively. The results of the standard forcing-based free-energy LB model are unavailable at $T/T_c = 0.7$ as the model is unstable in this case. The kinematic viscosity is chosen as $\nu = 0.15$. For comparison, the theoretical results are also presented in the figures. From Fig. 4 it can be seen that the liquid and gas densities predicted by the standard forcing-based free-energy LB model significantly deviate from those of the theoretical solution. In contrast, Figs. 4 and 5 show that the numerical results given by the improved scheme are in excellent agreement with the theoretical ones in both cases. Specifically, in the case of $T/T_c = 0.8$ the maximum relative errors yielded by the standard forcing-based free-energy LB model and the improved scheme are about 35.3% and 0.23%, respectively. Obviously, the free-energy LB model using the improved scheme performs much better than the standard forcing-based free-energy LB model.

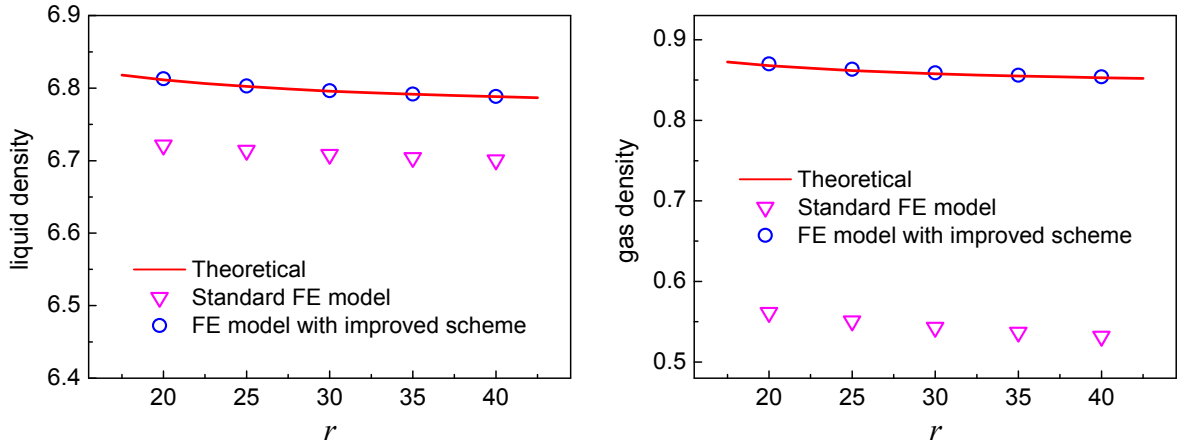

**FIG. 4**. Simulation of circular droplets. Comparison of the numerical liquid and gas densities obtained by different forcing-based free-energy LB models with the theoretical ones at $T = 0.8T_c$.

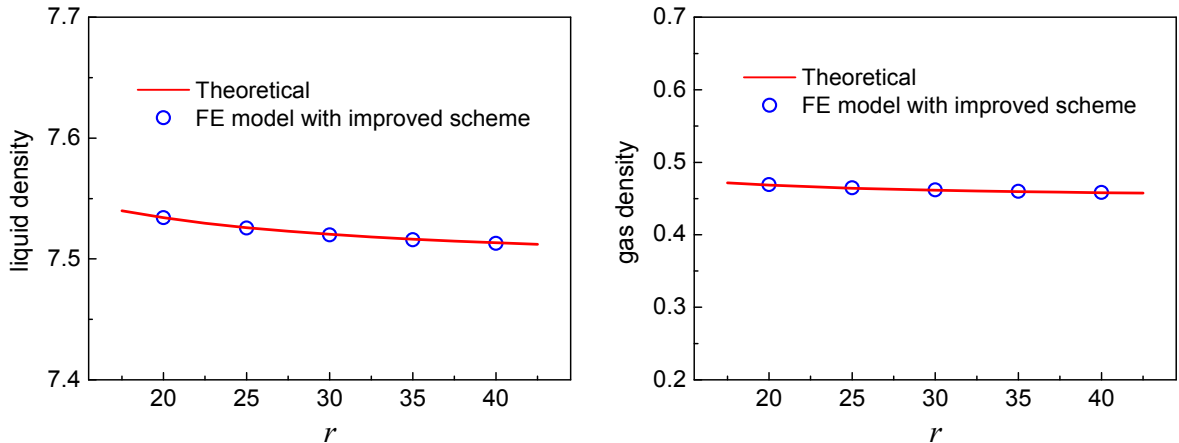

**FIG. 5**. Simulation of circular droplets. Comparison of the numerical liquid and gas densities obtained by the free-energy LB model using the improved scheme with the theoretical ones at $T = 0.7T_c$.

**Table I**. Comparison of the maximum spurious currents yielded by the standard forcing-based free-energy LB model and the free-energy LB model using the improved scheme in the simulation of circular droplets at $T = 0.8T_c$.

| Cases | $r = 20$ | $r = 25$ | $r = 30$ | $r = 40$ |
|---|---|---|---|---|
| Standard FE model | $5.56 \times 10^{-4}$ | $5.52 \times 10^{-4}$ | $5.42 \times 10^{-4}$ | $5.13 \times 10^{-4}$ |
| Improved scheme | $2.93 \times 10^{-15}$ | $3.32 \times 10^{-15}$ | $1.86 \times 10^{-15}$ | $2.20 \times 10^{-14}$ |

Table I provides a comparison of the maximum spurious currents produced by the standard forcing-based free-energy LB model and the free-energy LB model using the improved scheme in the simulation of circular droplets at $T = 0.8T_c$ with different initial radii. The results are obtained after $5 \times 10^4$ time steps. From the table we can see that in all the cases the maximum spurious currents yielded by the improved scheme are on the order of $10^{-14} \sim 10^{-15}$, which are smaller by ten orders of magnitude than those caused by the standard forcing-based free-energy LB model. The results demonstrate that the improved scheme is capable of not only eliminating the thermodynamic inconsistency but also significantly reducing the spurious currents.

As discussed in the previous section, when Eq. (25) is implemented without other changes, the LB model will suffer from the lost of Galilean invariance. In LB literature [14,26] it has been reported that a circular droplet in a uniform flow field will become an elliptic one when employing a two-phase LB model with broken Galilean invariance. To verify the Galilean invariance of the improved scheme, a moving circular droplet in a two-dimensional channel is simulated. The grid system is chosen as $L_x \times L_y = 120 \times 120$ and a circular droplet of $r = 25$ is initially placed at the center of the computational domain. The parallel top and bottom plates in the $y$ direction begin to move with a constant velocity $U = 0.1$ at $t = 0$. The no-slip boundary scheme [57] is applied at the solid plates and the periodic boundary condition is employed in the $x$ direction. The reduced temperature and the kinematic viscosity are taken as $T/T_c = 0.7$ and $\nu = 0.15$, respectively. Figure 6 displays some snapshots of a moving circular droplet simulated by the free-energy LB model with the improved scheme. It can be seen that the circular shape of the droplet is well preserved during the moving process, confirming that the improved scheme is Galilean invariant.

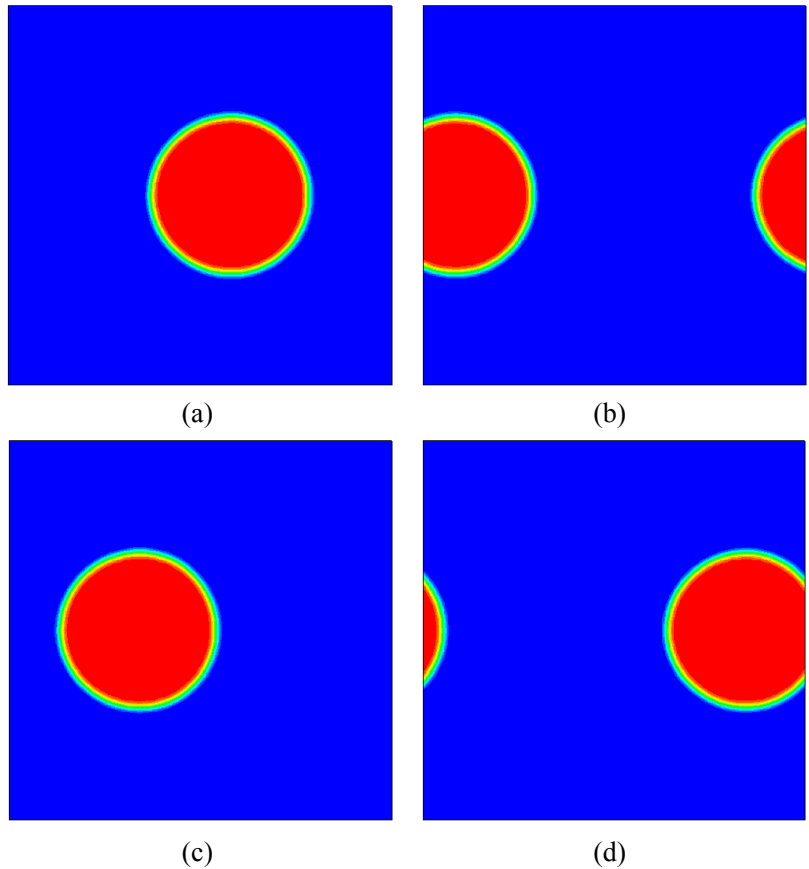

**FIG. 6**. Density contours of a moving circular droplet simulated by the free-energy LB model with the improved scheme. (a) $t = 20000\delta_t$, (b) $40000\delta_t$, (c) $70000\delta_t$, and (d) $90000\delta_t$.

## V. Summary

In this paper, we have investigated the problem of thermodynamic inconsistency of a class of free-energy LB models in which the divergence of thermodynamic pressure tensor or its equivalent form expressed by the chemical potential is incorporated into the LB equation via a forcing term. It is shown that the numerical error term that causes the thermodynamic inconsistency mainly includes two parts, one comes from the discrete gradient operator and the other can be identified in a high-order

Chapman-Enskog analysis. An improved scheme that is still within the framework of the standard streaming-collision procedure has been proposed to eliminate the thermodynamic inconsistency of the forcing-based free-energy LB models. The improved scheme is constructed by modifying the equation of state of the standard LB equation, through which the discretization of $\nabla(\rho c_s^2)$ is no longer required in the force calculation.

Numerical simulations have been performed for one-dimensional flat interfaces and two-dimensional circular droplets to validate the proposed scheme. It has been shown that the improved scheme is capable of eliminating the thermodynamic inconsistency and can significantly reduce the spurious currents in comparison with the standard forcing-based free-energy LB model. Specifically, in the test of circular droplets, the maximum spurious currents yielded by the improved scheme are on the order of $10^{-14} \sim 10^{-15}$, which are smaller by ten orders of magnitude than those produced by the standard forcing-based free-energy LB model. Finally, it should be mentioned that using the chemical potential is not the only way to incorporate the surface tension effect. Some other ways can be found in Ref. [58], in which a comparative analysis of various surface tension formulations has been made.

## Acknowledgments

This work was supported by the National Natural Science Foundation of China (No. 51822606).**Appendix A: Comparison of the improved scheme with two available schemes in the literature**

In this appendix, the improved scheme devised in the present work is numerically compared with the Lax-Wendroff propagation scheme formulated by Lou and Guo [32] and the forcing scheme proposed by Wagner [49] (e.g., see Eq. (59) in Ref. [49]) for eliminating the thermodynamic inconsistency of the forcing-based free-energy LB models. The Lax-Wendroff propagation scheme [32] can be expressed as follows:

$$f_\alpha^*(\mathbf{x},t) = f_\alpha(\mathbf{x},t) - \frac{1}{\tau}\left[f_\alpha(\mathbf{x},t) - f_\alpha^{eq}(\mathbf{x},t)\right] + \hat{\delta}_t\left(1 - \frac{1}{2\tau}\right)G_\alpha, \quad (A1)$$

$$f_\alpha\left(\mathbf{x}, t+\hat{\delta}_t\right) = a_0 f_\alpha^*(\mathbf{x}, t) + a_1 f_\alpha^*\left(\mathbf{x}+\mathbf{c}_\alpha \delta_x, t\right) + a_{-1} f_\alpha^*\left(\mathbf{x}-\mathbf{c}_\alpha \delta_x, t\right), \tag{A2}$$

where $a_0 = 1 - A^2$, $a_1 = A(A-1)/2$, $a_{-1} = A(A+1)/2$, and $\mathbf{c}_\alpha = \mathbf{e}_\alpha/c$, in which $A$ is the Courant-Friedrichs-Lewy (CFL) number and $c$ is the lattice constant. In lattice units, the spatial step $\delta_x$ and the lattice constant are usually taken as $\delta_x = c = 1$. But the time step $\hat{\delta}_t$ in Eqs. (A1) and (A2) is determined from the CFL condition [32], i.e., $\hat{\delta}_t = A\delta_x/c$, which is different from the time step $\delta_t$ in the standard LB method.

According to Ref. [32], the numerical errors can be reduced by decreasing the CFL number. A typical choice of the CFL number is $A = 0.1$, which leads to $\hat{\delta}_t = 0.1\delta_t$. Correspondingly, the computational cost will be increased by ten times. For the above Lax-Wendroff propagation scheme, the macroscopic velocity is calculated via $\rho\mathbf{u} = \sum_\alpha \mathbf{e}_\alpha f_\alpha + 0.5\hat{\delta}_t \mathbf{F}$, in which $\mathbf{F} = \nabla(\rho c_s^2) - \rho\nabla\mu_c$, and the equilibrium density distribution function in Eq. (A1) is chosen as [32]

$$f_\alpha^{eq} = \rho\omega_\alpha\left[1 + \frac{(\mathbf{e}_\alpha \cdot \mathbf{u})}{c_s^2} + \frac{(\mathbf{e}_\alpha \cdot \mathbf{u})^2}{2c_s^4} - \frac{|\mathbf{u}|^2}{2c_s^2} + \frac{B\hat{\delta}_t\left(\mathbf{e}_\alpha\mathbf{e}_\alpha - c_s^2\mathbf{I}\right):\mathbf{T}}{2c_s^2}\right], \tag{A3}$$

where $B$ is a free parameter used to adjust the kinematic viscosity and the tensor $\mathbf{T}$ is calculated by

$$\rho c_s^2 \hat{\delta}_t (B-\tau)\mathbf{T} = \sum_\alpha \mathbf{e}_\alpha \mathbf{e}_\alpha f_\alpha - \rho c_s^2 \mathbf{I} - \rho\mathbf{u}\mathbf{u} - (\tau - 0.5)\hat{\delta}_t(\mathbf{u}\mathbf{F} + \mathbf{F}\mathbf{u}). \tag{A4}$$

The kinematic viscosity is then defined as $\nu = c_s^2(\tau - 0.5 - B)\hat{\delta}_t$.

The test of flat interfaces is examined in this appendix. The simulation setup, the grid system, and the parameters of the van der Waals equation of state are the same as those used in Sec. IV. According to Ref. [32], the CFL number of the Lax-Wendroff propagation scheme is taken as $A = 0.1$. The parameter $B$ in Eq. (A3) is determined from $\nu = c_s^2(\tau - 0.5 - B)\hat{\delta}_t$ with $\tau = 1$ and $\hat{\delta}_t = A = 0.1$. In simulations, the kinematic viscosity is chosen as $\nu = 0.04$. Therefore, for the Lax-Wendroff propagation scheme the parameter $B$ is given by $B = -0.7$ and the relaxation time is taken as $\tau = 0.62$ for the forcing scheme proposed in Ref. [49]. The coexisting gas and liquid densities obtained by the Lax-Wendroff propagation scheme, the forcing scheme proposed in Ref. [49], and the improved scheme devised in the present work are compared in Table II for the cases of $T/T_c = 0.9$, 0.85, 0.8, and 0.7, respectively. For comparison, the analytical results of the Maxwell construction are also listed there.

From the table it can be seen that the numerical results obtained by the improved scheme are in excellent agreement with the analytical ones in all the investigated cases, whereas the gas density given by the Lax-Wendroff propagation scheme gradually deviates from the analytical solution when the reduced temperature $T/T_c$ decreases and the maximum relative error is about 3.5%, which is produced at $T/T_c = 0.7$ in the gas phase. The numerical results of the forcing scheme proposed in Ref. [49] agree well with the analytical results of the Maxwell construction in the cases of $T/T_c = 0.9$ and 0.85, but it suffers from numerical instability at $T/T_c = 0.8$ and 0.7.

**Table II**. Comparison of the coexisting gas and liquid densities obtained by the Lax-Wendroff scheme [32], the forcing scheme proposed in Ref. [49], and the improved scheme at $\nu = 0.04$.

| $T/T_c$ | L-W scheme [32] | | Scheme in Ref. [49] | | Improved scheme | | Analytical | |
|---|---|---|---|---|---|---|---|---|
| | $\rho_g$ | $\rho_l$ | $\rho_g$ | $\rho_l$ | $\rho_g$ | $\rho_l$ | $\rho_g$ | $\rho_l$ |
| 0.9 | 1.4865 | 5.7992 | 1.4901 | 5.80045 | 1.4901 | 5.80045 | 1.4901 | 5.80045 |
| 0.85 | 1.113 | 6.323 | 1.11905 | 6.325 | 1.11905 | 6.325 | 1.11905 | 6.32499 |
| 0.8 | 0.8300 | 6.7626 | — | — | 0.83883 | 6.76447 | 0.83883 | 6.76447 |
| 0.7 | 0.4324 | 7.4896 | — | — | 0.44805 | 7.49149 | 0.44805 | 7.4915 |

The chemical potential profiles predicted by the Lax-Wendroff propagation scheme, the forcing scheme proposed in Ref. [49], and the improved scheme are compared in Fig. 7, from which it can be found that the improved scheme produces a uniform chemical potential in all the investigated cases, while the Lax-Wendroff propagation scheme leads to a fluctuating chemical potential with variations across the interfaces. Specifically, the numerical chemical potentials given by the improved scheme agree very well the analytical results obtained from Eq. (30), which are given by $\mu_c^{an} \approx 0.041974$, $0.030243$, $0.018302$, and $-0.006307$ for the cases of $T/T_c = 0.9$, 0.85, 0.8, and 0.7, respectively. The results of the forcing scheme proposed in Ref. [49] are unavailable at $T/T_c = 0.8$ and 0.7 owing to numerical instability, but it produces a uniform chemical potential in the cases of $T/T_c = 0.9$ and 0.85. The different performances between the Lax-Wendroff propagation scheme and the improved scheme may be attributed to the fact that the discretization of $\nabla(\rho c_s^2)$ is involved in the former scheme but is not needed in the latter scheme.

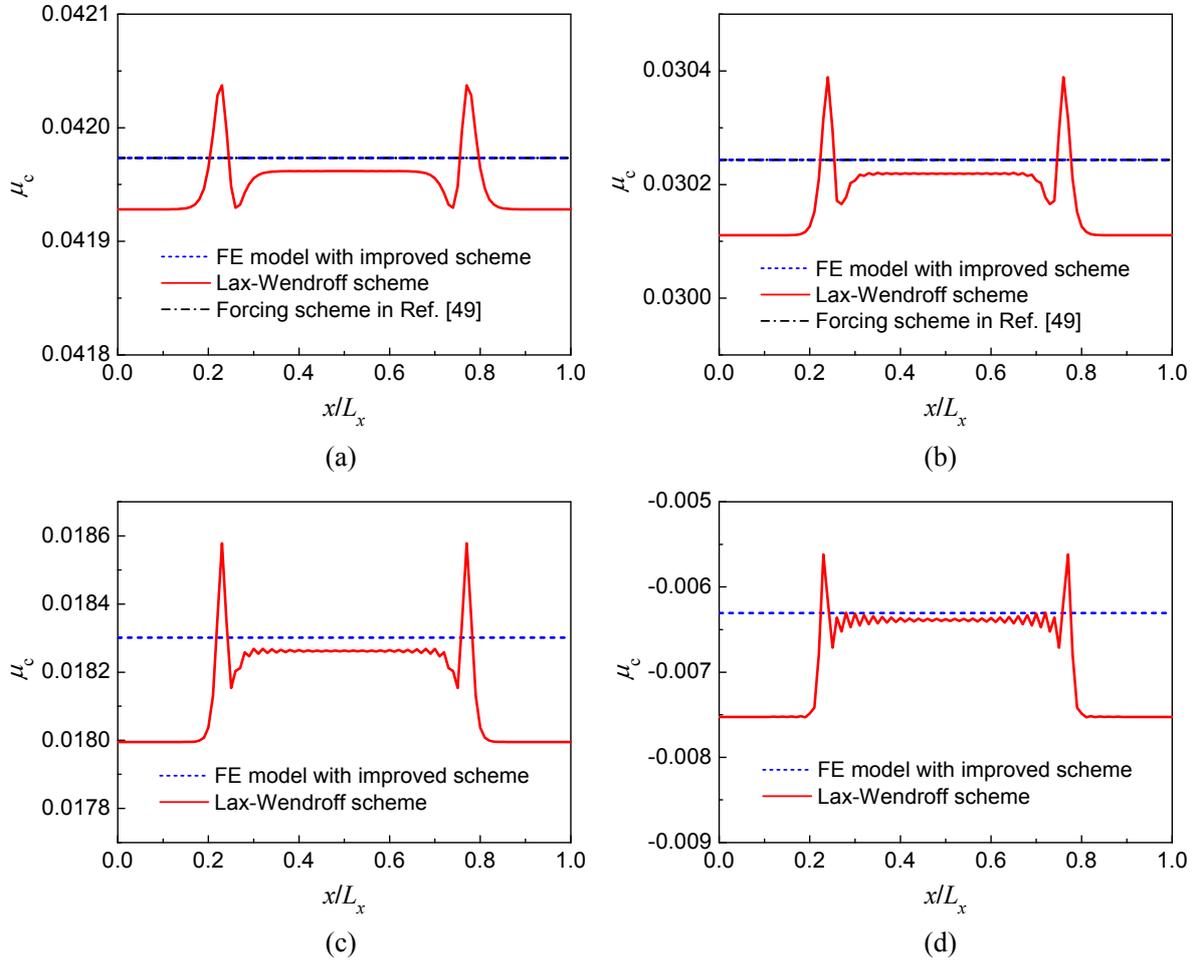

**FIG. 7**. Comparison of the chemical potential profiles predicted by the Lax-Wendroff scheme [32], the forcing scheme proposed in Ref. [49], and the improved scheme devised in the present work at $\nu = 0.04$. (a) $T = 0.9T_c$, (b) $T = 0.85T_c$, (c) $T = 0.8T_c$, and (d) $T = 0.7T_c$.